\documentclass[3p,twocolumn]{elsarticle}
\usepackage{amsmath,amssymb,amsfonts}
\usepackage{subcaption}
\usepackage{algorithmic}
\usepackage{graphicx}
\usepackage{textcomp}
\usepackage{xcolor}
\usepackage{tablefootnote}
\usepackage{hyperref}
\usepackage{multirow}
\usepackage{pifont}
\newcommand{\xmark}{\ding{53}}%
\journal{Journal of \LaTeX\ Templates}

\begin{document}
	
	\begin{frontmatter}
		
		\title{Text-Conditioned Transformer for  Automatic Pronunciation Error Detection}

		%%\author{Elsevier\fnref{myfootnote}}
		%%\address{Radarweg 29, Amsterdam}
		%%\fntext[myfootnote]{Since 1880.}
		
		\author[mymainaddress]{Zhan Zhang}
		\ead{zhan\_zhang@zju.edu.cn}
		
		\author[mymainaddress]{Yuehai Wang\corref{mycorrespondingauthor}}
		\cortext[mycorrespondingauthor]{Corresponding author}
		\ead{wyuehai@zju.edu.cn}
		
		\author[mymainaddress]{Jianyi Yang}
		\ead{yangjy@zju.edu.cn}
		
		\address[mymainaddress]{ Department of Information and Electronic Engineering, Zhejiang University, China}

		\begin{abstract}
			Automatic pronunciation error detection (APED) plays an important role in the domain of language learning. As for the previous ASR-based APED methods, the decoded results need to be aligned with the target text so that the errors can be found out. However, since the decoding process and the alignment process are independent, the prior knowledge about the target text is not fully utilized. In this paper, we propose to use the target text as an extra condition for the Transformer backbone to handle the APED task. The proposed method can output the error states with consideration of the relationship between the input speech and the target text in a fully end-to-end fashion.
			Meanwhile, as the prior target text is used as a condition for the decoder input, the Transformer works in a feed-forward manner instead of autoregressive in the inference stage, which can significantly boost the speed in the actual deployment.
			We set the ASR-based Transformer as the baseline APED model and conduct several experiments on the L2-Arctic dataset. The results demonstrate that our approach can obtain 8.4\% relative improvement on the $F_1$ score metric.
		\end{abstract}
		
		\begin{keyword}
			automatic pronunciation error detection (APED), computer-assisted pronunciation training (CAPT), Transformer
		\end{keyword}
		
	\end{frontmatter}
	
%	\linenumbers

	\section{Introduction}
	With the quick development of globalization and education, the number of language learners is rapidly increasing. However, most learners are facing the problem of teacher shortage or finding a proper time to follow systematic learning. Thus, recently, the computer-assisted language learning (CALL)\cite{beatty2013teaching} systems have been studied to offer a flexible education service, which can be used to reach the language learning requirement in fragmented time. In particular, oral practice is an important part of daily communication, and computer-assisted pronunciation training (CAPT)\cite{stenson1992effectiveness} systems are designed for this task. Such systems generally play the role of automatic pronunciation error detection (APED). The APED system first gives a predefined utterance text (and a reference speech of a professional teacher if needed), and the learner tries to pronounce this target text correctly. For example, a learner wants to study the pronunciation of ``apple'' (its phonemes are ``AE P AH L''), but the learner may mispronounce it to ``AE P AO L''. We call ``AE P AO L'' as the canonical pronunciation.
	By accurately detecting the pronunciation errors and providing precise feedback that ``AH'' is mispronounced, the APED system guides the learner to correct the pronunciation towards the target utterance and improve the speaking ability. 
	
	APED has been widely studied for decades. Depending on how to evaluate the matching degree between the student pronounced speech and the standard pronunciation, several comparison-based or goodness of pronunciation (GOP) methods have been proposed to solve the APED task\cite{lee2013pronunciation, lee2013mispronunciation, lee2016personalized, lee2012comparison,  witt1999use, witt2000phone}. 
	Recently, with the rising trend for neural networks and the development of automatic speech recognition (ASR) technologies, some end-to-end APED models \cite{leung2019cnn, zhang2020end} have been studied to simplify the workflow. They use ASR backbones to recognize the canonical pronunciation and obtain where the errors are, based on the alignment between the predicted phonemes and the standard phonemes. The ASR-based methods can significantly decrease the deploying efforts compared with conventional GOP methods or comparison-based methods. In particular, recently, the Transformer structure\cite{vaswani2017attention} shows a good performance for sequence-to-sequence (seq2seq) modelling, and gets promising performance in ASR tasks \cite{moritz2020streaming, zhang2020transformer, watanabe2018espnet, dong2018speech}. Thus, we choose the Transformer as the backbone for APED tasks in this paper.
	
	However,  the main deficiency of the {conventional ASR-based} Transformer for APED tasks is that the autoregressive decoding will slow the inference speed\cite{gu2017non}. Unfortunately, the APED task generally requires the system to give a quick response about the errors so that the learners can adapt their pronunciations and evaluate again. 
	Another consideration is that, for the ASR-based APED, the decoded text sequence needs to be aligned with the target text to detect the errors. Since the target text is already known in advance, it is a waste to ignore this prior knowledge during the autoregressive inference.  On the one hand, the length of the target text is fixed, but the autoregressive decoding is  length-agnostic. On the other hand, the recognized sequence is generally close to the prior target text in this evaluation task. These two factors inspire us to use the target text as extra input for the network.	
	
	In this paper, we propose an {ASR and alignment unified} Transformer-based APED workflow, which can incorporate both the audio feature and the text information, and output the error states directly. Compared with ASR-based methods which optimize the recognition result to improve the APED performance, the proposed method works in a fully end-to-end manner. Thus, the proposed method can optimize the APED metric directly. We observe a 8.4\% relative improvement on the $F_1$ score for the L2-Arctic dataset\cite{zhao2018l2} with the proposed method. Meanwhile, by using the prior target text as an input condition, the inference process works in a feed-forward manner rather than autoregressive, which can significantly boost the inference speed as suggested in \cite{ren2019fastspeech, peng2019parallel}. 
	
	The rest of this paper is organized as follows. In Section \ref{sec:related}, we analyze the related works about the APED task and how we are inspired to propose the text-conditioned {feed-forward} Transformer; In Section \ref{sec:method}, we compare the baseline { ASR-based autoregressive APED Transformer} and describe the {proposed ASR and alignment unified  feed-forward Transformer} in detail; Next, we analyze the results obtained by the conventional methods and the proposed method in Section \ref{sec:experiment}; Finally, we show the conclusion of this paper in Section \ref{sec:conclusion}.	
		
	\section{Related Works}
	\label{sec:related}
	From the perspective of language learning, an error detected in the APED system can be described as that the produced pronunciation is a nonstandard one. In other words, the pronounced speech deviates too far from the standard target speech. Based on this simple idea, comparison-based APED methods \cite{lee2013pronunciation, lee2013mispronunciation, lee2016personalized, lee2012comparison} have been explored. These methods generally adopt dynamic time warping (DTW) \cite{berndt1994using} algorithms to align the extracted features of the input speech with the standard target speech. Depending on the distance between each text unit, the pronunciation quality score can be calculated. To this end, the comparison-based methods need to prepare a standard speech for reference, which are inconvenient to evaluate a new utterance. 
	
	Apart from directly comparing to a specific standard speech, the input speech can also be evaluated by whether a standard acoustic model can recognize each phoneme. In particular, the likelihood of each phoneme has proven to be an effective feature for indicating whether the error happens, and such a likelihood-based scoring method is often referred to as GOP \cite{witt1999use, witt2000phone}. In practice, this approach utilizes the hidden Markov model (HMM) to model the sequential phone states. The likelihood score is calculated from the force-aligned states and the open phone states. Since the first proposal of GOP by  \cite{witt1999use}, many variants \cite{kim1997automatic, franco1999automatic, proencca2017detection, hu2013new, cheng2015deep} have been studied to adapt its original equation for better measurement of the goodness.
	
	With the rise of deep learning, the performance of the ASR tasks has been greatly improved. Thus, by utilizing the advanced acoustic model of an ASR system and recognizing the input speech, ASR-based APED can be another efficient approach to detect the errors. Such a method can also avoid the deploying efforts of conventional HMM-based GOP methods or comparison-based DTW methods, and several ASR-based APED systems have been proposed  \cite{leung2019cnn, zhang2020end}. Currently, the ASR systems are generally built upon CTC loss 
	\cite{graves2006connectionist} or attention mechanism \cite{jan2015attention,chan2016listen} to handle the sequential features. The main deficiency of CTC loss is the conditional independent assumption. Such an assumption may not be valid for the continuous speech. The ASR performance is reported to be better by combining the CTC loss with the attention mechanism \cite{watanabe2017hybrid} or using the Transformer structure\cite{watanabe2018espnet, dong2018speech}.  In particular, the Transformer structure, which is originally designed to handle the natural language processing (NLP) problems \cite{dai2019transformer, devlin2018bert}, has been successfully utilized in several other domains, such as computer vision (CV)\cite{ribeiro2020sketchformer, carion2020end}, and speech-related tasks including text to speech (TTS) \cite{okamoto2020transformer, li2019neural, ren2019fastspeech, peng2019parallel}, voice conversion (VC)\cite{liu2020voice}, and ASR \cite{moritz2020streaming, zhang2020transformer}.
	
	Despite the convenience of ASR-based APED systems, alignment is still an inevitable process to obtain the final evaluation results. The recognized phonemes should be aligned with the target phonemes to find out the mispronunciations. As the alignment process is not integrated into the backward optimization of the ASR model, such a method is not fully end-to-end.	In other words, the decoding process and the evaluation process are independent.
	However, intuitively, human raters will first keep the target text in mind, then try to compare the input speech to find out where the errors take place. Focussing on the prior target text limits the search space for the decoding process. Extended Recognition Network (ERN) \cite{harrison2009implementation} utilizes this idea to incorporate prior knowledge about common mispronunciations into the HMM states. However, the predefined error HMM paths will lead to bad performance when faced with unseen mispronunciations. Despite its weakness, ERN still shows that the prior knowledge is of vital importance to facilitate the performance of APED tasks. This inspires us to directly take the prior target text as an extra condition, together with the speech features for input. Meanwhile, the attention mechanism can be a logical approach to fuse both the speech feature and the text feature. Thus, the attention-based seq2seq models including Listen, attend and spell (LAS)\cite{chan2016listen} and Transformer\cite{vaswani2017attention} are ideal backbones to start with. Transformer uses the positional encoding to model the time information, instead of a recurrent architecture in LAS. The ASR performance of Transformer is reported to be better in \cite{dong2018speech}. Thus, we use Transformer as the backbone in this paper.
	
	However, the {conventional} attention-based Transformer generally adopt autoregressive decoding to predict the next entity. This will lead to a slow inference, which can be a deficiency for the APED system. As analyzed in \cite{gu2017non}, for each decoding step, the current prediction depends on the earlier decoded output to get the conditional probability. However, 
	since the output target is already known in the training stage, the Transformer can assume this target as a decoded result (this is called as ``teacher-forcing''). Thus, the Transformer do not need to wait for the decoded output and the Transformer can run in parallel. In contrast, this prior does not exist in the inference stage, and the Transformer must run sequentially to predict the next entity for several decoding steps until meeting the end-of-sentence-tag ($\langle$EOS$\rangle$). {On the contrary, Transformers which work in a feed-forward manner can greatly boost the speed  \cite{gu2017non,ren2019fastspeech, peng2019parallel}.} Thus, for the APED task, {if we can utilize the prior text to be evaluated, and unify the ASR and alignment process, the conventional Transformers can decode in a feed-forward manner,} and the aforementioned limitation will no longer exist.
	
	Based on the analysis above, we propose the text-conditioned {ASR and alignment unified feed-forward} Transformer for the APED task. We give a detailed description of the proposed method in the next section.

	\section{Proposed Method}
	\label{sec:method}
	In this section, we first show the conventional ASR-based APED workflow for comparison. Next, we demonstrate the proposed fully end-to-end workflow and describe the network structure and its training method in detail.
	\subsection{ASR-Based APED}
	\label{ssec:asr}
	\begin{figure}[!htp]%htb	
		\begin{minipage}[b]{1.0\linewidth}
			\centering
			\centerline{\includegraphics[width=1\linewidth]{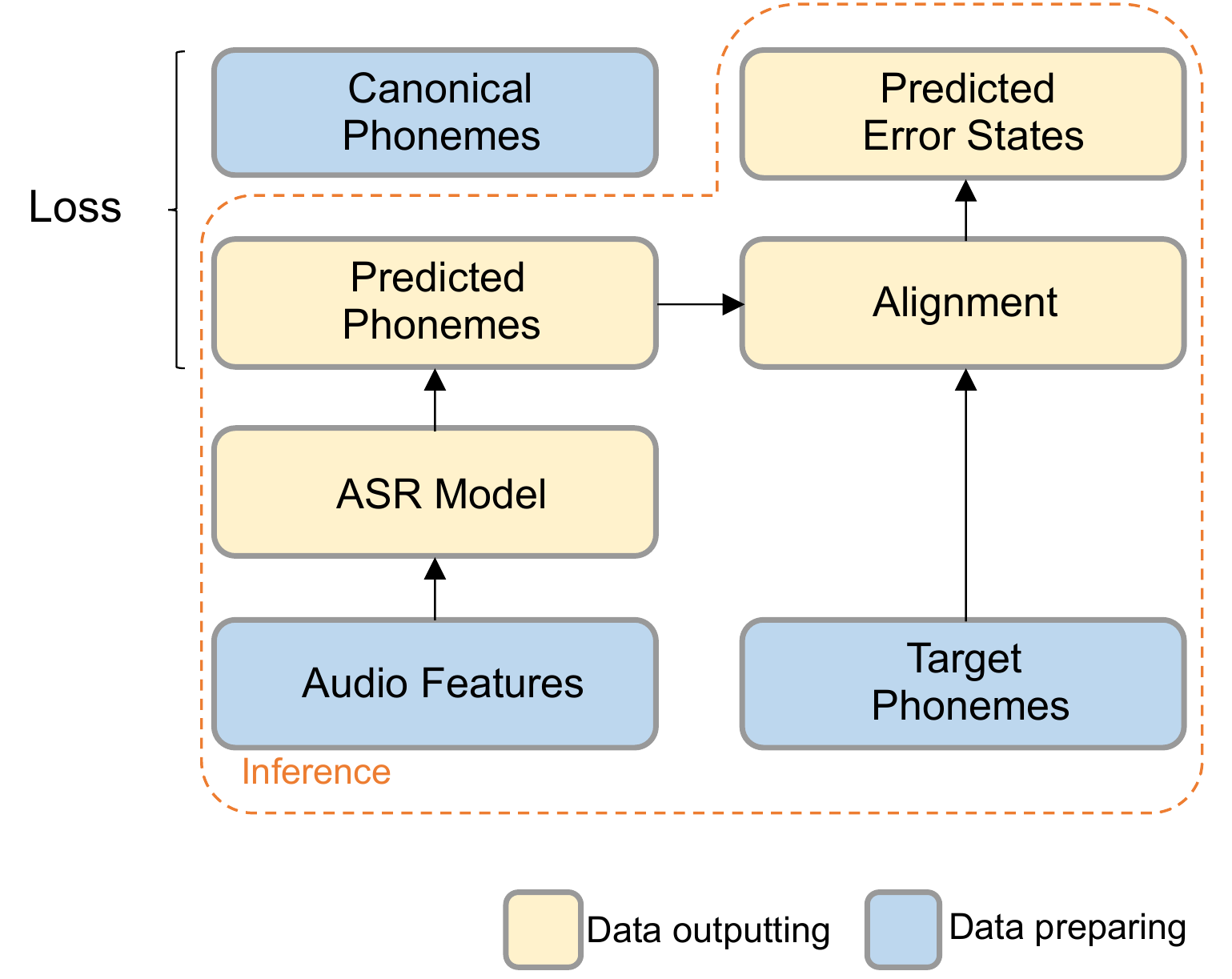}}		
		\end{minipage}
		\caption{Workflow for the ASR-based APED method. The alignment process is independent from the phoneme prediction process.}
		\label{fig:workflow0}
	\end{figure}
	
	A typical workflow for the ASR-based APED is depicted in Fig.\ref{fig:workflow0}. The training dataset is generally constructed by three parts, the target text to be read, the collected speech, and the canonical pronounced text marked by professional annotators. For example, L2-Arctic dataset\cite{zhao2018l2} manually labels the correct phonemes and mispronunciation error tags about the collected speech. Three  annotators who are experienced in transcribing speech samples of native or non-native English speakers participate in the annotating process to ensure the high quality. Based on such a dataset, an ASR model is trained to recognize the canonical phoneme-level text $\mathbf{p}=(p_1,p_2,...,p_n, p_{n+1})$ from the extracted audio features $\mathbf{x}=(x_1,x_2,...,x_m)$.
	We should note that the described {ASR-based APED} is general, and can be applied with any ASR systems that can translate the audio features into phonemes. However, as we focus on Transformer, we limit our description on the attention-based training and inference in the following paragraph.
	For the attention-based models, the cross-entropy loss is used between the predict phonemes $\hat{\mathbf{p}}$ and the canonical phonemes $\mathbf{p}$:
	\begin{equation}
	l_{asr}=CrossEntropy(\hat{\mathbf{p}}, \mathbf{p}),
	\end{equation}
	where $p_{n+1}= \langle$EOS$\rangle$.

	For the inference stage, the Transformer works quite differently from the training stage. The Transformer uses autoregressive outputting method to recognize the canonical phonemes sequentially. The recognized phonemes string will end with $\langle$EOS$\rangle$. Next, Needleman-Wunscha algorithm\cite{likic2008needleman} is applied to align the recognized sequence $\hat{\mathbf{p}}$ with the target phonemes $\mathbf{t}=(t_1,t_2,...,t_k)$. After the alignment process, the error states $\mathbf{e}=(e_1,e_2,...,e_k)$ with consideration of the target phonemes can be returned to the user. An alignment example is shown in Table \ref{tab:sample}.  We can observe that this sample includes 1 deletion and 2 substitution errors. The mispronounced phonemes whose error states are marked as 1 can be returned to the users.
	
	\begin{table*}[htbp]
		\centering
		\caption{Alignment sample}		
		\resizebox{\textwidth}{!}{
			\begin{tabular}{ccccccccccccccccccccccc}
				\hline
				\textbf{} & \multicolumn{2}{c}{IF} & \multicolumn{2}{c}{YOU} & \multicolumn{4}{c}{ONLY} & \multicolumn{3}{c}{COULD} & \multicolumn{2}{c}{KNOW} & \multicolumn{2}{c}{HOW} & I & \multicolumn{4}{c}{THANK} & \multicolumn{2}{c}{YOU} \\
				\textbf{Target} & IH & F & Y & UW & OW & N & L & IY & K & UH & D & N & OW & HH & AW & AY & TH & AE & NG & K & Y & UW \\
				\textbf{Pronounced} & IH & F & Y & UW & {\color[HTML]{FE0000} AO} & N & L & IY & K & UH & {\color[HTML]{FE0000} -} & N & {\color[HTML]{FE0000} AO} & HH & AW & AY & TH & AE & NG & K & Y & UW \\
				\textbf{Error States} & 0 & 0 & 0 & 0 & {\color[HTML]{FE0000} 1} & 0 & 0 & 0 & 0 & 0 & {\color[HTML]{FE0000} 1} & 0 & {\color[HTML]{FE0000} 1} & 0 & 0 & 0 & 0 & 0 & 0 & 0 & 0 & 0 \\ \hline
			\end{tabular}	
		}
		\label{tab:sample}
	\end{table*}
	
	For better clarification, we summarize the training and the inference stage of the ASR-based model in Table \ref{tab:asrti}. We use a 39-dim Mel frequency cepstral coefficients (MFCC) feature as the encoder input. The start-of-sentence tag ($\langle$SOS$\rangle$) and the right-shifted 1-dim label of the canonical phonemes are concatenated as the decoder input in the training stage. This input is replaced by $\langle$SOS$\rangle$ and a regressively decoded phonemes string in the inference stage. The decoder tries to predict the probability of the next phoneme and $\langle$EOS$\rangle$ for output. There are in total 42 tags for classification, including 39 phonemes and $\langle$SOS$\rangle$ $\langle$EOS$\rangle$ $\langle$PAD$\rangle$.
	
	\begin{table*}[]
		\centering
		\caption{Training and inference summary of the ASR-Based Transformer}
		\resizebox{0.7\textwidth}{!}{
			\begin{tabular}{cccc}
				\textbf{} & \multicolumn{3}{c}{\textbf{Training Stage}} \\
				\hline
				\textbf{} & \textbf{EncoderInput} & \textbf{DecoderInput} & \textbf{DecoderOutput} \\
				\textbf{data}& SpeechFeatures & $\langle$SOS$\rangle$+Canonical Phonemes(Shifted) & Canonical Phonemes+$\langle$EOS$\rangle$ \\
				\textbf{loss} &- &- & $l_{asr}$ \\
				\textbf{len} & m & 1+n & n+1 \\
				\textbf{dim} & 39 & 1 & 42 \\
				& \multicolumn{3}{c}{\textbf{Inference Stage}} \\
				\hline
				& \textbf{EncoderInput} & \textbf{DecoderInput} & \textbf{DecoderOutput} \\
				\textbf{data}& SpeechFeatures & $\langle$SOS$\rangle$+Recognized Phonemes & Next Recognized Phonemes \\
				\textbf{len} & m & End with $\langle$EOS$\rangle$ & End with $\langle$EOS$\rangle$ \\
				\textbf{dim} & 39 & 1 & 42
			\end{tabular}
		}
		\label{tab:asrti}
	\end{table*}
	
	We should note that there are several lengths defined for the described sequences. First, the attention mechanism is adopted to match the speech features ($length=m$) and the recognized phonemes ($length=n+1$). Next, the alignment operation is applied to find out the error states, whose length is equal to that of the target phonemes ($length=k$). 
	However, such an alignment operation is performed in the inference stage,  thus not jointly optimized with the ASR model. Such a dilemma inspires us to integrate the alignment operation or the target text into the training stage.

	\subsection{Fully End-to-end APED}
	\begin{figure*}[!htp]%htb
		\centering
		\begin{minipage}[b]{0.7\linewidth}
			\centering
			\centerline{\includegraphics[width=1\linewidth]{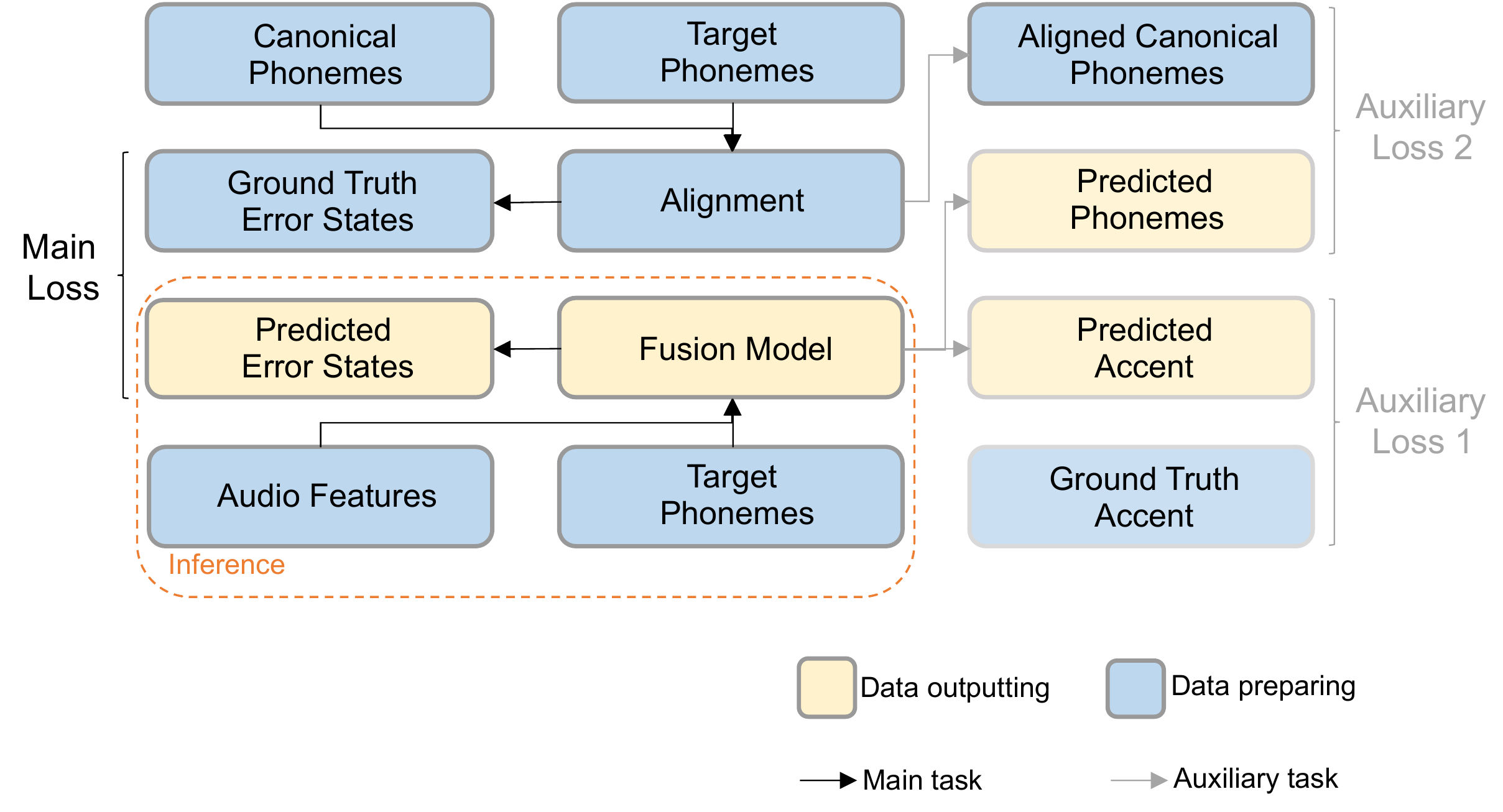}}		
		\end{minipage}
		\caption{Workflow for the proposed APED method. We move the alignment process into the preparing stage. The proposed model can directly output the error states. Meanwhile, the auxiliary accent and phoneme classification tasks are adopted.}
		\label{fig:workflow1}
	\end{figure*}
	As shown in Fig.\ref{fig:workflow1}, for the proposed method, we move the alignment operation into the data preparing stage. We align the canonical phonemes and the target phonemes to obtain where the errors occur in advance.   
	\begin{figure}[htp!]%htb	
		\begin{minipage}[b]{1.0\linewidth}
			\centering
			\centerline{\includegraphics[width=1\linewidth]{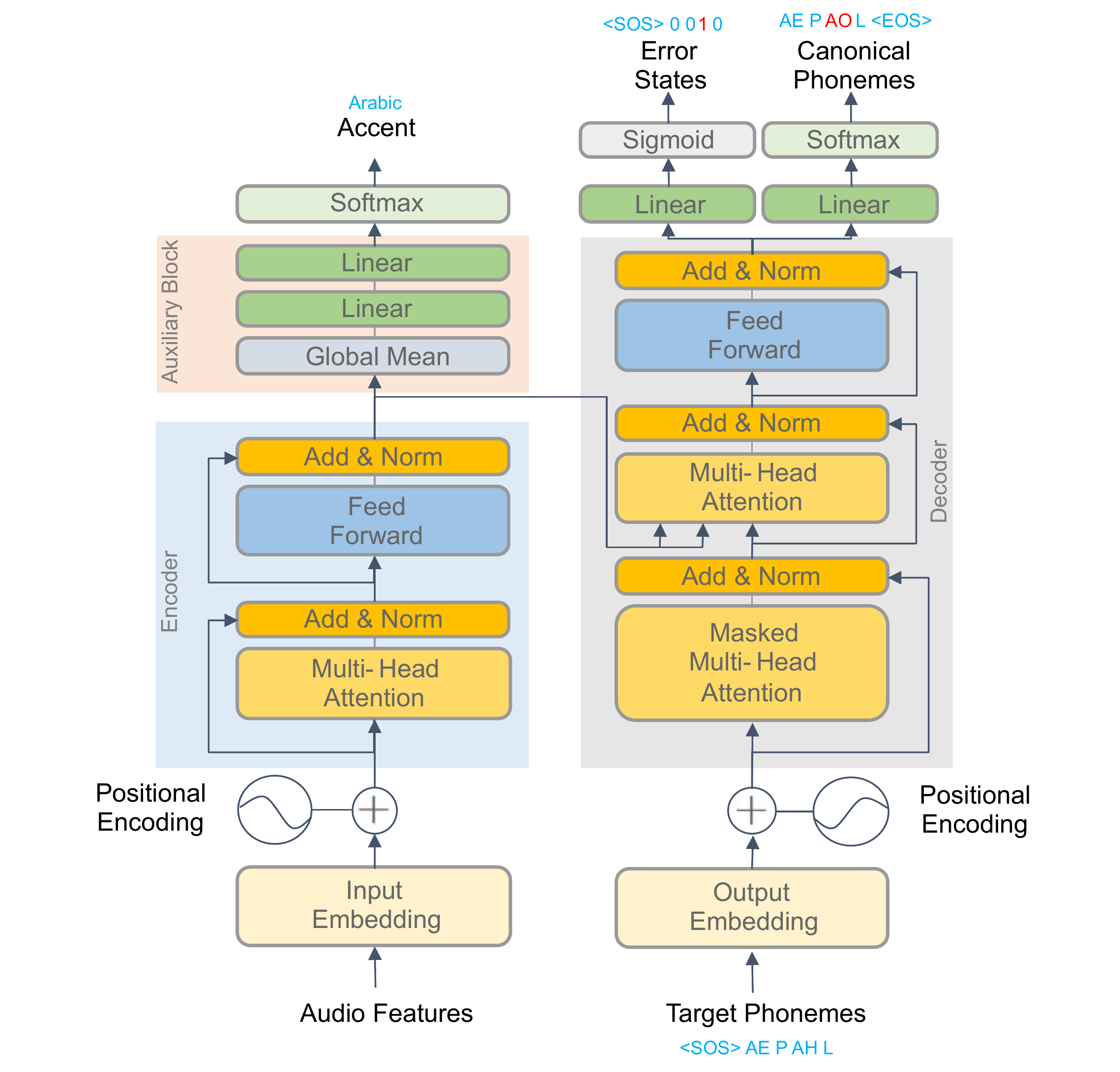}}		
		\end{minipage}
		\caption{Network architecture of the text-conditioned Transformer. We append an accent classifier after the encoder to extract the L1-related information. Target phonemes are used as an extra condition for the decoder input. The error states are obtained in a feed-forward manner. Meanwhile, phoneme classification is also performed as an auxiliary task. The mispronounced word ``APPLE" is shown in this figure for demonstration.}
		\label{fig:model1}
	\end{figure}
	
	Next, we directly evaluate the relationship between speech features and the target phonemes. Thus, the network can be viewed as a fusion model. 	
	Moreover, the mother language (L1) of the speakers shows to affect the acoustic characteristics when studying a new language (L2) \cite{chang2010first, jiao2016accent}. Meanwhile, the extracted L1 features have also been proved to be helpful in the APED task \cite{tu2018investigating}. Thus, we introduce the accent related auxiliary task to extract the L1  information. As shown in Fig.\ref{fig:model1}, we append an extra classifier after the encoder and use the cross-entropy loss between the predicted accent $\hat{a}$ and the ground truth accent $a$ presented in the dataset:
	\begin{equation}
	l_{a}=CrossEntropy(\hat{a}, a).
	\end{equation}

	Since the speech evaluation dataset is scarce, we first obtain a basic acoustic model by training the model on ASR datasets. The training process is similar to conventional ASR-based APED methods discussed in Section \ref{ssec:asr}, and the new ASR loss function is,
	\begin{equation}
	l_{asr}^{'}=l_{asr}+\alpha l_{a},
	\label{eq:asr}
	\end{equation}
	where $\alpha$ is the weight of the auxiliary accent task.
	
	We further adapt this basic acoustic model to the APED task. A training and inference summary of the proposed model is shown in Table \ref{tab:ourti}. We will discuss the details and the differences between the proposed model and the ASR-based model in the remaining paragraphs.
	\begin{table*}[htp!]
		\caption{Training and inference summary of the proposed Transformer}
		\centering
		\resizebox{1\textwidth}{!}{
			\begin{tabular}{cccccc}
				\textbf{} & \multicolumn{5}{c}{\textbf{Training Stage}} \\
				\hline
				\textbf{} & \textbf{EncoderInput} & \textbf{EncoderOutput} & \textbf{DecoderInput} & \textbf{DecoderOutput1} & \textbf{DecoderOutput2} \\
				\textbf{data}& SpeechFeatures & Accent & $\langle$SOS$\rangle$+Target Phonemes & Aligned Canonical   Phonemes+$\langle$EOS$\rangle$ & $\langle$SOS$\rangle$+Error States \\
				\textbf{loss} &- & $l_a$ &- & $l_{asr}$ & $l_{eval}$ \\
				\textbf{len} & m & 1 & 1+k & k+1 & 1+k \\
				\textbf{dim} & 39 & 6 & 1 & 42 & 1 \\
				& \multicolumn{5}{c}{\textbf{Inference Stage}} \\
				\hline
				& \textbf{EncoderInput} & \textbf{EncoderOutput} & \textbf{DecoderInput} & \textbf{DecoderOutput1} & \textbf{DecoderOutput2} \\
				\textbf{data}& SpeechFeatures & Accent & $\langle$SOS$\rangle$+Target Phonemes & Canonical Phonemes+$\langle$EOS$\rangle$ & $\langle$SOS$\rangle$+Error States \\
				\textbf{len} & m & 1 & 1+k & k+1 & 1+k \\
				\textbf{dim} & 39 & 6 & 1 & 42 & 1
			\end{tabular}
		}
		\label{tab:ourti}
	\end{table*}
	
	Firstly, for the auxiliary accent classification task, while the input audio features are sequential, the accent is a 1-dim global attribute. We try to process the sequential data with gated recurrent units  (GRU)\cite{cho2014learning} or a simple GlobalMean. Experiments in Section \ref{sec:experiment} show that GlobalMean performs a little better. Note that there are 6 kinds of accent including Arabic, Chinese, Hindi, Korean, Spanish, and Vietnamese, for the used dataset in our experiments.
	
	Secondly, the prior target phonemes are used as an extra condition for the decoder input instead of the canonical pronounced phonemes, in both the training and the inference stage. For the audio features $\mathbf{x}$ and a certain target phoneme $t_i$ (target phoneme at step $i$), the decoder output is changed to $\hat{e_i}$, which indicates the matching degree of the audio features and $t_i$. As we use a binary state to judge its goodness, we use the sigmoid activation at the last layer for binary classification in Fig.\ref{fig:model1}.
	
	As the whole process is differentiable, we can directly optimize the loss between the predicted error states $\hat{\mathbf{e}}$ and the ground truth error states $\mathbf{e}$. For now, several classification losses can be used for this model. We first apply a basic binary cross-entropy (BCE) loss between the predicted error states $\hat{\mathbf{e}}=(\hat{e_0},\hat{e_1},\hat{e_2},...,\hat{e_k})$ and the ground truth error states $\mathbf{e}=(e_0, e_1,e_2,...,e_k)$ as the evaluation loss,
	\begin{equation}
	l_{eval}^{BCE}=BCE(\hat{\mathbf{e}},\mathbf{e}),
	\label{eq:bce}
	\end{equation}
	where $e_0=\langle$SOS$\rangle$.
	A further discussion about the choice of loss functions is presented in Section \ref{ssec:loss}.
	
	However, compared with ASR-based methods, a binary state only concerns about whether the target phoneme is correct or mispronounced. Thus, the model may lose information about the exact phoneme. To fix this, we still require the proposed model to conduct the ASR task with an auxiliary weight of $\beta$, and the whole loss function is,
	\begin{equation}
	l=l_{eval}+\beta l_{asr}+\alpha l_a.
	\label{eq:proposed}
	\end{equation}
	The canonical phonemes to be recognized are aligned with the target phonemes for the proposed model using the aforementioned Needleman-Wunscha algorithm in Section \ref{ssec:asr} to make these two phoneme strings have equal length $k+1$.
	
	Lastly, we should note that the proposed model has a consistent behavior in the training and inference stage, as shown in Table \ref{tab:ourti}. This characteristic makes the inference in our method faster compared with ASR-based autoregressive Transformers, shown in Section \ref{ssec:latency}.

	\section{Experiment}
	\label{sec:experiment}
	We use the SpeechTransformer backbone proposed in \cite{li2019speechtransformer} for experiments. The SpeechTransformer is constructed by 6 encoder and 6 decoder layers in our experiments. Meanwhile, the attention modelling dimension $d_{model}=512$, 4 attention heads, and the feed-forward dimension $d_{ff}=1024$ are adopted. We extract the MFCC features of the audio files by Kaldi toolkit\cite{povey2011kaldi}. These MFCC features are subsampled with a factor of $n=4$, and stacked with $m=5$ number of frames, which is the same as the settings in \cite{li2019speechtransformer}.
	We demonstrate the ASR performance for phoneme recognition in the first subsection \ref{ssec:phoneme}. Then we use this pretrained model to adapt for the APED task and show the latency, metric, and results in the next three subsections. Finally, we analyze the loss functions and the behavior of the proposed model in the last two subsections, \ref{ssec:loss} and \ref{ssec:analysis}, correspondingly.
	
	\subsection{Phoneme Recognition}%Comparison with previous models
	\label{ssec:phoneme} 
	We use Librispeech  \cite{panayotov2015librispeech} as the dataset for ASR training.  This dataset contains approximately 1000 hours of 16kHz read English speech. It is divided into ``clean'' (460 hours) and ``other'' (500 hours) parts based on its recognition difficulty. The ``clean'' part is further divided into the training set of 100 hours and 360 hours, development set (dev-clean), and test set (test-clean). As the APED task focuses on the phoneme-level error, we first convert the dataset into phoneme-level transcriptions using the Montreal Forced Aligner tool\cite{mcauliffe2017montreal}. Next, we train the Transformer on different parts of the trainset for 300 epochs, including train-clean of 100 hours (train-100h), the whole train-clean part (train-460h), and the whole train part (train-960h). We use dev-clean as the validation dataset to choose the best model and test-clean for inference performance comparison. Adam optimizer, with a learning rate of $10^{-3}$, is used. We use a CTC-based ASR model called Jasper5x3 proposed in \cite{li2019jasper} for comparison. This model is constructed by 5 repeated Jasper blocks, and each Jasper block is constructed by 3 repeated Conv1D sub-blocks. The model parameters of Jasper5x3 are 44M, while the proposed Transformer is 32M. We show the phone error rate (PER) performance in Table \ref{tab:asr}. As we can see from the table, {for different amounts of training resources,} the attention-based Transformer structure generally performs better than the CTC-based method on PER. This observation is in accord with the conclusion in \cite{watanabe2018espnet,dong2018speech}, as the attention mechanism in Transformer can capture more relevant information compared with the CTC loss which holds the conditional independent assumption.

	% \usepackage{booktabs}
	\begin{table}[htbp]
		\begin{center}
			\caption{Performance of PER {on different subsets of} Librispeech dataset.}
			\resizebox{0.45\textwidth}{!}{
				\begin{tabular}{@{}ccccc@{}}
					\textbf{} & \multicolumn{2}{c}{\textbf{CTC-Based}} & \multicolumn{2}{c}{\textbf{Transformer-Based}} \\
					train-        & dev-clean         & test-clean         & dev-clean\footnotemark             & test-clean             \\
					\hline
					100h           & 8.13\%            & 8.50\%             & 4.55\%                & 8.11\%                 \\
					460h           & 4.88\%            & 5.50\%             & 2.32\%                & 4.24\%                 \\
					960h           & 4.02\%            & 4.23\%             & 1.70\%                & 3.17\%                 \\
					\hline
					
				\end{tabular}
			}
			\label{tab:asr}
		\end{center}
	\end{table}
	\footnotetext[1]{Since this result on the development set dev-clean is obtained by using the teacher-forcing training, PER is much lower.}

	\subsection{Latency Experiment}
	\label{ssec:latency}
	Next, we conduct the APED task on  L2-Arctic dataset \cite{zhao2018l2}. This corpus contains 26,867 utterances with 6 different accents, from 24 nonnative speakers.  The 3,599 utterances annotated on phoneme-level are used for the APED task. The trainset, valset, and testset are divided into 8:1:1. {For the testset, each sentence contains about 30 target phonemes on average.} {This suggests that the conventional autoregressive ASR-based models need to forward about 30 times on average to get each decoded phoneme sequentially. On the contrary, the proposed methods only need to forward once.} We conduct the latency evaluation on a server with Intel Xeon E5-2680 CPU, and 1 NVIDIA P100 GPU. As shown in Table \ref{tab:latency}, the proposed method can bring great speedup for the APED inference.

	\begin{table}[!htp]
		\caption{Latency comparsion. $b$ stands for batchsize here. The latency is computed as the average time to decode each sentence in the testset.}
		\resizebox{0.45\textwidth}{!}{
\begin{tabular}{ccc}
	\multicolumn{1}{l}{} & \multicolumn{1}{l}{\textbf{Latency(ms)}} & \multicolumn{1}{l}{\textbf{Speedup}} \\ \hline
	ASR-Based ($b=1$) & 1194±198 & 1.00$\times$ \\
	ASR-Based ($b=4$) & 966±106 & 1.24$\times$ \\
	Proposed ($b=1$) & 88±13 & 13.6$\times$ \\
	Proposed ($b=4$) & 67±6 & 17.8$\times$
\end{tabular}
	}
\label{tab:latency}

	\end{table}

	\begin{table*}[!htb]
	\centering
	\caption{Comparison between different models.}
	\resizebox{\textwidth}{!}{
		\begin{tabular}{ccccccccc}
			& \begin{tabular}[c]{@{}c@{}}\textbf{Accent}\\      \textbf{Classification}\end{tabular} & \begin{tabular}[c]{@{}c@{}}\textbf{Phoneme}\\      \textbf{Classification}\end{tabular} & \textbf{FAR} & \textbf{FRR} & \textbf{Acc} & \textbf{Precision} & \textbf{Recall} & \textbf{F1} \\
			\hline
			\textbf{GOP-Based} & \textbf{} & \textbf{} &  &  &  &  &  &  \\
			GMM-HMM(Librispeech)\footnotemark &- &- & - & - & - & 0.290 & 0.290 & 0.290 \\
			\textbf{ASR-Based} & \textbf{} & \textbf{} &  &  &  &  &  &  \\
			Initial(Librispeech) &\xmark & \checkmark & 0.485 & 0.207 & 0.753 & 0.295 & 0.515 & 0.375 \\
			Fine-tuned(L2-Arctic)&\xmark  &\checkmark  & 0.375 & 0.103 & 0.858 & 0.504 & 0.625 & 0.558 \\
			Fine-tuned(L2-Arctic)& \checkmark &\checkmark  & 0.353 & 0.106 & 0.859 & 0.507 & 0.647 & 0.568 \\
			\textbf{Proposed} & \textbf{} & \textbf{} &  &  &  &  &  &  \\
			BCE Loss & \checkmark & \xmark & 0.458 & 0.051 & 0.890 & 0.639 & 0.542 & 0.587 \\
			BCE Loss& \checkmark & \checkmark & 0.429 & 0.054 & 0.890 & 0.641 & 0.571 & 0.603 \\
			F1 Loss & \checkmark & \xmark & 0.442 & 0.055 & 0.889 & 0.630 & 0.558 & 0.591 \\
			F1 Loss& \checkmark & \checkmark & 0.428 & 0.058 & 0.889 & 0.622 & 0.572 & 0.596 \\
			Focal Loss & \checkmark & \xmark & 0.424 & 0.060 & 0.888 & 0.617 & 0.576 & 0.595 \\
			Focal Loss& \checkmark & \checkmark & 0.423 & 0.055 & 0.882 & 0.636 & 0.577 & \textbf{0.605}\\
			\hline
		\end{tabular}
	}
	\label{tab:comparsion}
\end{table*}

	\subsection{APED Metric}
	\label{ssec:metric}
	For the APED task, the model should make a good balance of detecting the wrong pronunciations and accepting the correct ones. Thus, $F_1$ score is chosen as the main indicator for the performance. As defined in \cite{qian2010discriminative}, the hierarchical evaluation structure is first divided into correct pronunciations and wrong pronunciations by the canonical pronounced phoneme. Next, depending on whether the predicted error state matches the ground truth label, the outcomes are further divided into true acceptance (TA), false rejection (FR), false acceptance (FA), and true rejection (TR). In other words, T/F suggests whether the prediction of the model is correct for the APED task, and A/R is the decision of the model. Based on this evaluation structure, $F_1$ score of the APED system is defined as follows:
	\begin{equation}
	Precision=\frac{TR}{TR+FR},
	\end{equation}
	\begin{equation}
	Recall=\frac{TR}{TR+FA},
	\end{equation}
	\begin{equation}
	F_1=2\frac{Precision*Recall}{Precision+Recall}.
	\label{equ:f_1}
	\end{equation}
	{ To count $TR, TA, FR, FA$ for metrics,} the predicted binary error states $(\hat{e_1},\hat{e_2},...,\hat{e_k})$ are firstly filtered by a threshold of $\theta=0.5$ to transform from a {continuous} float with the range of $(0,1)$ into {discrete} binary integer $\{0, 1\}$,
	\begin{equation}
	\hat{e}\leftarrow
	\begin{cases}
	1,& \text{if } \hat{e}\geq \theta\\
	0.              & \text{otherwise}
	\end{cases}
	\label{eq:filter}
	\end{equation}
	Next, each outcome is calculated by following equations,
	\begin{equation}
	TR=\sum_{i=1}^{k}(\hat{e_i}*e_i),
	\label{eq:tr}
	\end{equation}
	\begin{equation}
	FR=\sum_{i=1}^{k}(\hat{e_i}*(1-e_i)),
	\label{eq:fr}
	\end{equation}	
	\begin{equation}
	FA=\sum_{i=1}^{k}((1-\hat{e_i})*e_i),
	\label{eq:fa}
	\end{equation}	
	\begin{equation}
	TA=\sum_{i=1}^{k}((1-\hat{e_i})*(1-e_i)).
	\label{eq:ta}
	\end{equation}

	Apart from the conventional classification-related metrics including $F_1$ score, accuracy, precision and recall, the false rejection rate (FRR) and the false acceptance rate (FAR) are also of vital importance to the APED task. They are calculated as follows,
	\begin{equation}
	FRR=\frac{FR}{TA+FR},
	\end{equation}
	\begin{equation}
	FAR=\frac{FA}{FA+TR}.
	\end{equation}
	
	\footnotetext[2]{This result is taken from \cite{zhao2018l2}, Fig.4. It is trained on Librispeech train-960 and tested on L2-Arctic dataset.}	 
	
	\subsection{APED Result}
	\label{ssec:aped}
\begin{figure}[htb!]%htb	
		\begin{minipage}[b]{1.0\linewidth}
			\centering
			\centerline{\includegraphics[width=1\linewidth]{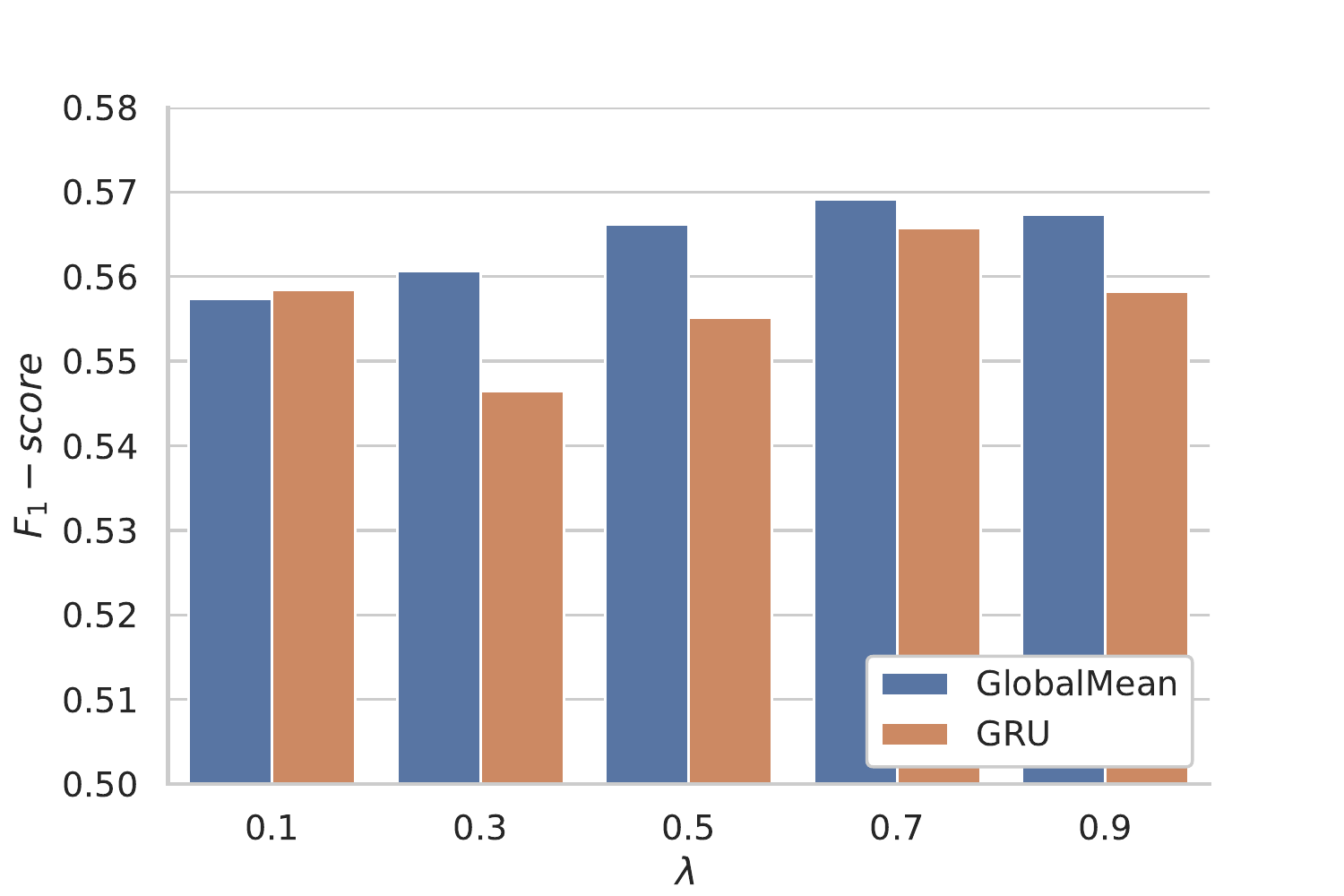}}		
		\end{minipage}
		\caption{$F_1$ score comparison of GlobalMean and GRU.}
		\label{fig:auxiliary}
	\end{figure}

	We first conduct experiments to explore the auxiliary accent classification task. We start from the model obtained on Librispeech dataset, and train for another 200 epochs, with the learning rate decreased to $10^{-4}$.  We find that the GlobalMean method performs a little better than the GRU, as shown in Fig.\ref{fig:auxiliary}. We use $\alpha=0.7$ for the ASR-based Transformer in Eq.\ref{eq:asr}, and lower it to $\alpha=0.1$ in Eq.\ref{eq:proposed} to balance the $l_{asr}$ loss for further experiments about the proposed text-conditioned version.
	
	% \usepackage{multirow}

	Next, we adapt this pretrained ASR-based model to the proposed text-conditioned version. We still train the whole model for 200 epochs, with the learning rate of $10^{-4}$. 
	We set the ASR-based model without the auxiliary task for baseline and the proposed methods with ablation for comparison. We find that $\beta=0.3$ generally gives the best performance.
	The results are shown in Table \ref{tab:comparsion}. The performance of the GOP method tested on this dataset\cite{zhao2018l2} and the initial Transformer model pretrained on Librispeech is also reported in this table. First of all, as the initial model is purely trained on Librispeech, which only includes standard pronunciations,  its FRR is relative high as it directly treats some unseen accents as wrong pronunciations. When adapted to handle the L2-Arctic dataset, the model has a significant improvement. By further employing the accent auxiliary task, the $F_1$ score is increased by nearly 0.1. If we simply use the target text as the condition and change the prediction target to the error states,  the basic binary cross-entropy loss can bring a 0.19 improvement in terms of the $F_1$ score. 
	We discuss the effect of different loss functions for the proposed method in the next subsection.

	\subsection{Loss functions}
	\label{ssec:loss}
	First of all, as the $F_1$ score is an important metric for the APED task, inspired by \cite{eban2017scalable}, we directly utilize the generalized $F_1$ score to optimize the predicted error states.
	To make it differentiable, the sums of probabilities are used instead of counts. We do not apply Eq.\ref{eq:filter}  before calculating Eq.\ref{eq:tr} - \ref{eq:ta}. {We should also note that Eq.\ref{eq:filter} is  applied only for metrics calculation. For all the loss functions discussed in this subsection, $\hat{e}$ is continuous.} As we try to maximize the $F_1$ score, the $F_1$ evaluation loss of the proposed method is,
	\begin{equation}
	l_{eval}^{F_1}=1-F_1.
	\label{equ:eval_loss}
	\end{equation}

	Another consideration is that only 14.56\% of the labelled phone segments are mispronounced for the L2-Arctic dataset\footnote{Dataset document at \url{https://psi.engr.tamu.edu/l2-arctic-corpus-docs/}}, which may cause an unbalance between correct pronunciations and mispronunciations. 
	Thus, we adopt focal loss\cite{lin2017focal} to mine the hard labels.
	Formally, if we define $e_t$ as:
	\begin{equation}
	e_t=
	\begin{cases}
	\hat{e},& \text{if }  e=1\\
	1-\hat{e}.              & \text{otherwise}
	\end{cases}
	\end{equation}
	The focal loss is,
	\begin{equation}
	l_{eval}^{focal}=-(1-e_t)^\gamma log(e_{t}),
	\end{equation}
	where $\gamma$ modulates how much the well-classified samples are down-weighted. When $\gamma=0$, this loss function is equivalent to Eq.\ref{eq:bce}.
	
	We apply $F_1$ loss function and focal loss with different $\gamma$ values to the proposed model. We can see from Table \ref{tab:comparsion} that, when adopting the $F_1$ loss function instead of the basic BCE loss, the result can be slightly improved. For the focal loss, we find that a small $\gamma$ value ($\gamma$=0.5 in our experiments) performs the best, and a bigger value will lead to a degraded $F_1$ score. Meanwhile, the auxiliary ASR task can boost the performance for all these loss functions. The focal loss version has the highest $F_1$ score 0.605 for the default $\theta=0.5$, which is a relative 8.4\% improvement over the baseline ASR-based method.

	\subsection{Analysis}
	\label{ssec:analysis}
	We further analyse the behavior of the proposed method. 
	
	For the APED task, we need to make a trade-off between FAR and FRR. Meanwhile, as noted in \cite{eskenazi2009overview}, it is usually more unacceptable to take the correct pronunciations as wrong ones (false reject) than to regard the mispronunciations as correct ones (false acceptance).  We can observe from Table \ref{tab:comparsion} that the proposed methods all have a higher FAR and decreased FRR compared with ASR-based models, which suggests our model is behaving in a more acceptable way. 
	
\begin{figure*}[!htb]
	\centering
	\begin{minipage}[b]{0.4\linewidth}
	\includegraphics[width=\linewidth]{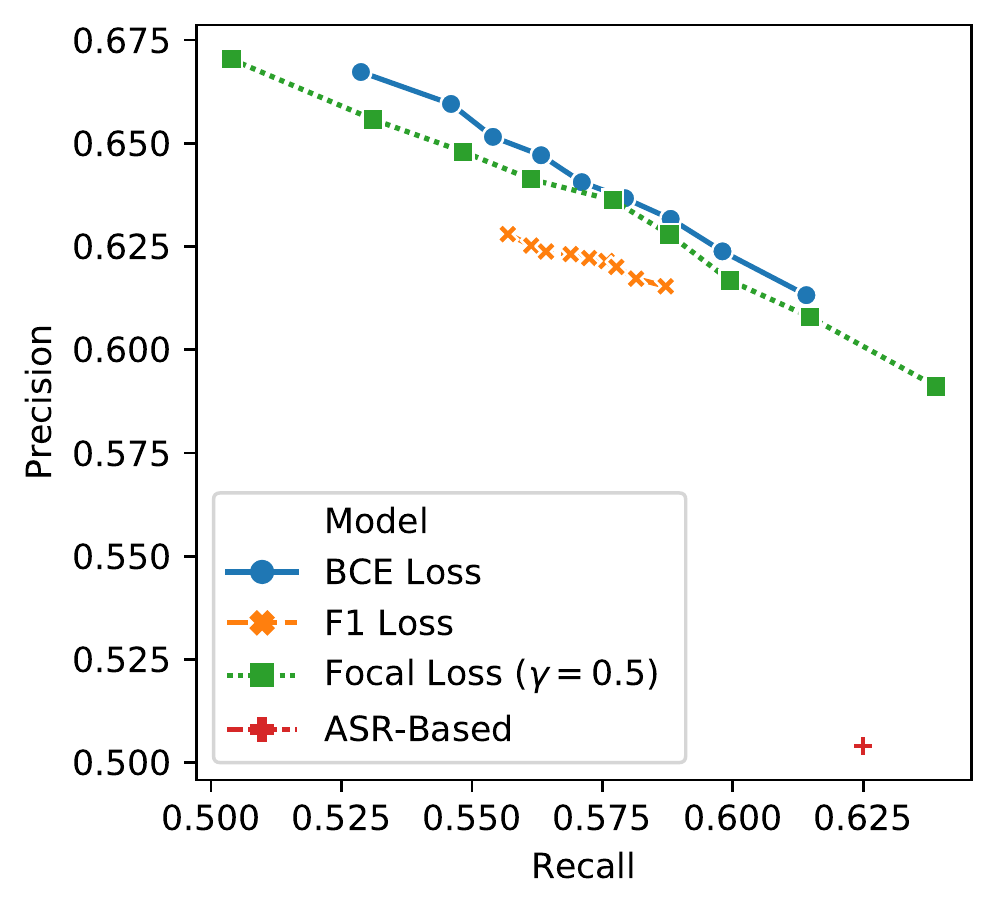}
	\subcaption{}
	\label{fig:prc}
	\end{minipage}
	\begin{minipage}[b]{0.4\linewidth}
	\includegraphics[width=\linewidth]{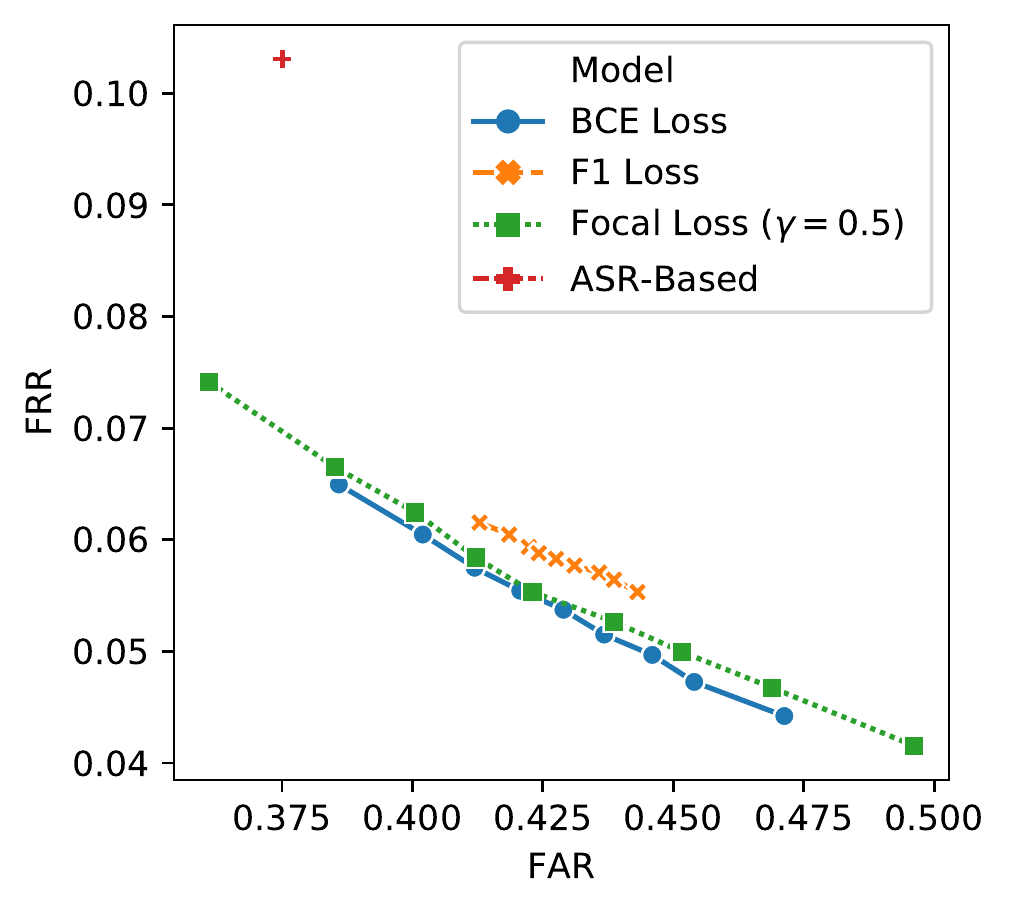}
	\subcaption{}
	\label{fig:arc}
	\end{minipage}
	
	\caption{Metrics between different models by changing $\theta$. We also show the performance of the ASR-based baseline. (a) Recall-Precision curves. (b) FAR-FRR curves.}
	\label{fig:changing_theta}
	
\end{figure*}

	For the actual deployment, as the proficiency level of the target language varies among different students, the trade-off between FAR and FRR should be easy to adjust. Compared with ASR-based models, the proposed method can simply change the threshold $\theta$ to control how strict the APED system is. We further explore the effect of changing $\theta$ for different loss functions. We use a step of 0.1, i.e., $\theta\in[0.1, 0.2, ..., 0.9]$. The metrics are shown in Fig.\ref{fig:changing_theta}. By increasing $\theta$, according to Eq.\ref{eq:filter}, more output is judged as correct, and less output is judged as error. As a result, FAR  (and precision) increases, while FRR (and recall) drops. Compared with the F1 loss version, the BCE loss and the focal loss version have a wider range of FAR and FRR when adjusting $\theta$ and can be a better choice for the actual deployment.

\begin{table*}[!htb]
	\centering
	\begin{tabular}{ccccc}
		\textbf{Quantile} & \textbf{Error Rate} & \textbf{ASR-Based F1} & \textbf{Proposed F1} & \textbf{Improvement} \\ \hline
		25\% & {[}0,7.8\%{]} & 0.338 & 0.390 & 15.38\% \\
		50\% & (7.8\%,13.0\%{]} & 0.461 & 0.516 & 11.93\% \\
		75\% & (13.0\%,19.4\%{]} & 0.550 & 0.609 & 10.73\% \\
		100\% & (19.4\%,46.8\%{]} & 0.674 & 0.694 & 2.97\%
	\end{tabular}
	\caption{{Pronunciation} error rate breakdown on the testset.}
	\label{tab:err}
\end{table*}

	\begin{figure*}[!htb]%htb
	\centering
	\begin{minipage}[b]{\linewidth}
		\centering
		\includegraphics[width=\linewidth]{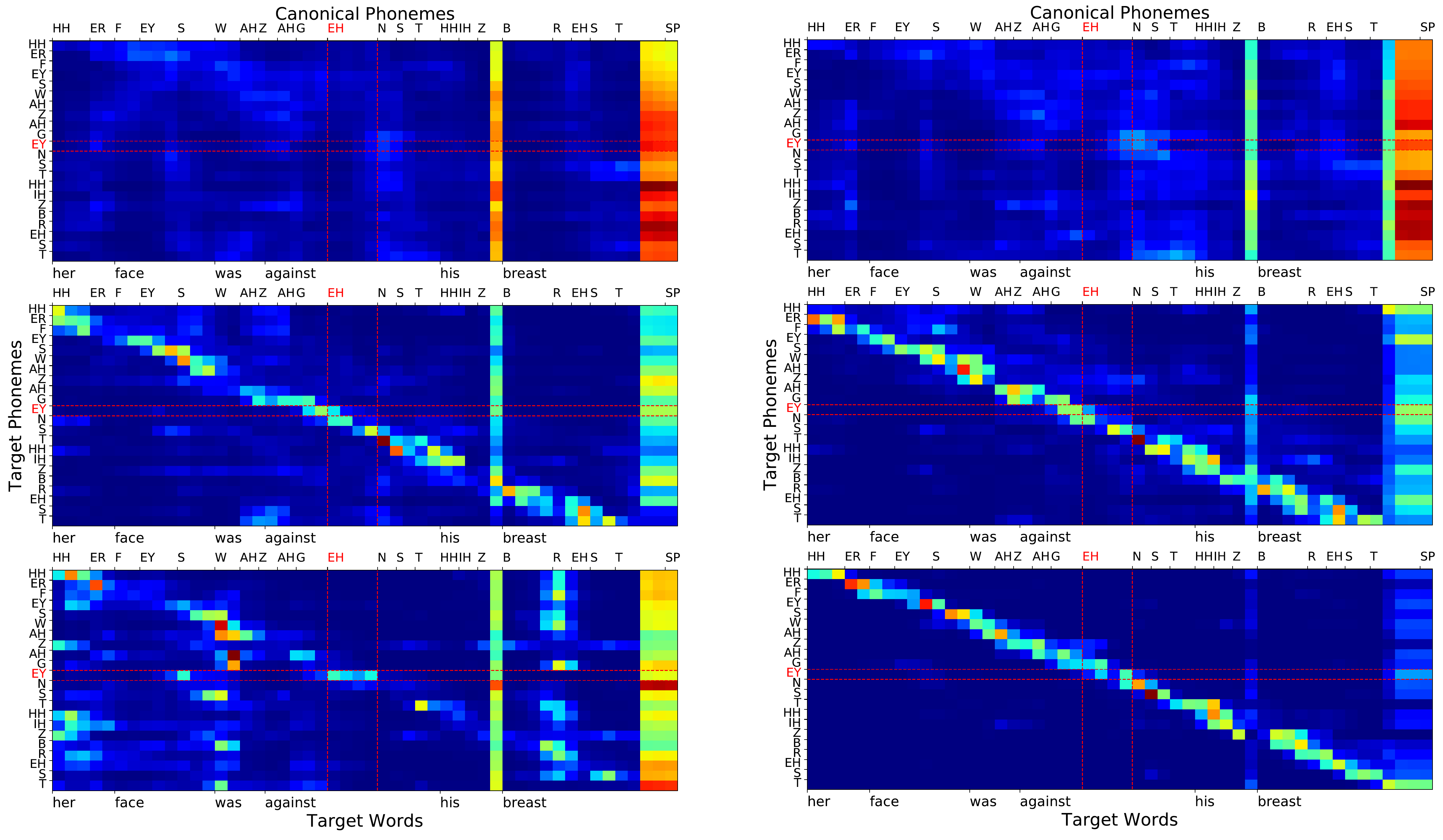}
	\end{minipage}

	\caption{Comparison between the simple version without ASR auxiliary task (left) and the full version (right). This sample (arctic\_a0129) is spoken by speaker ZHAA, ``HER FACE WAS AGAINST HIS BREAST”. The EY phoneme in ``AGAINST” is pronounced to be EH by mistake.
	When an error happens, the corresponding phoneme is marked as red. The simple version tries to do the alignment only for the shallow layers, whereas the behavior of the full version is more to the ASR-based Transformers.  }
	\label{fig:attention_asr}
\end{figure*}
	Since the proposed method is related to the input text, we break the testset into different parts {to show the impact of how much the canonical phonemes differ from the target phonemes, i.e., the pronunciation error rate.} A larger {pronunciation} error rate suggests that the input text information becomes less related to the input speech. We use the focal loss version and the ASR-based baseline for comparison. The result is shown in Table \ref{tab:err}. {We can observe that the proposed method makes higher relative improvement when the error rate is lower.}
	 
	Finally, we try to analyze the auxiliary ASR task. For the ASR-based Transformer, on the one hand, the encoder extracts the speech-related features as embeddings; On the other hand, the decoder uses the attention mechanism to query the corresponding weight of each memory for the input text. Thus, the attention mechanism in the decoder will conduct an alignment between the text and the corresponding speech. What will the proposed text-conditioned Transformer do if the input text is not the canonical pronunciation but the target one? We plot the attention map of the proposed method without or with the auxiliary ASR task in Fig.\ref{fig:attention_asr} to explore its behavior. For simplicity, we call these two models as the simple version and the full version in the following discussion.

	As shown in Fig.\ref{fig:attention_asr}, for the simple version, it still tries to align the speech feature with the target phonemes for the shallow layers. As for deeper layers, the alignment between the phonemes and the speech features becomes vague. We conjecture that the training target causes this phenomenon. As for the ASR task, the network has to predict the next phoneme exactly; However, for the APED task, the network just needs to handle the error pattern for each phoneme and outputs a binary state, which is an easier task. Under such a cosy target, the deeper layers may not work hard to do the alignment, but choose to focus on summarizing the error patterns.
	As pretraining on the ASR task can be viewed as a sequential adaption\cite{mao2020survey}, the pretrained weights perform as a regularization for the APED optimization space. Meanwhile, as suggested in \cite{sanh2020movement}, the adapted model does not deviate from the pretrained weights significantly. Thus, based on the pretrained ASR weight, the model still has the ability to distinguish different phonemes and match the input phonemes with the audio features memory. This may be the reason that the simple version can still get a satisfying improvement, as shown in   \ref{tab:comparsion}. When the model is required to conduct ASR task, namely, the full version, the attention maps appear to be regular, which are similar to those Transformers that are applied for ASR tasks. As a result, the full version generally performs better than the simple version.

	\section{Conclusion}
	\label{sec:conclusion}

	In this study, we propose a text-conditioned Transformer for automatic pronunciation error detection. By conditioning the target phonemes as an extra input, the Transformer can directly evaluate the relationship between the input speech and the target phonemes. Thus, the error states are obtained in a fully end-to-end manner. Meanwhile, unlike the conventional autoregressive Transformer, the proposed method works in a feed-forward manner in both the training and the inference stage.	We conduct a number of experiments to compare the performance of different methods and find that the proposed text-conditioned Transformer can boost the $F_1$ score of the APED task on the L2-Arctic dataset. The proposed method has a more reasonable FAR and FRR, and the degree of strictness can be easily adjusted by the threshold $\theta$ parameter.

	\bibliography{refs}
	
\end{document}